\documentclass{article}
\pdfoutput=1
\usepackage{textcomp}
\usepackage[square,numbers]{natbib}
\usepackage{hyperref}
\usepackage{amsmath}
\usepackage{amssymb}
\usepackage{stmaryrd}
\usepackage{mathtools}
\usepackage{bbold}
\usepackage{mathpartir}
\usepackage{verbatim}
\usepackage{booktabs}
\usepackage{placeins}
\usepackage[
    type={CC},
    modifier={by},
    version={4.0},
    imagewidth={8em},
]{doclicense}

\newtheorem{theorem}{Theorem}[section]

\DeclareSymbolFont{sfoperators}{OT1}{cmss}{m}{n}
\DeclareSymbolFontAlphabet{\mathsf}{sfoperators}

\makeatletter
\def\operator@font{\mathgroup\symsfoperators}
\DeclareFontFamily{OMX}{MnSymbolE}{}
\DeclareSymbolFont{MnLargeSymbols}{OMX}{MnSymbolE}{m}{n}
\SetSymbolFont{MnLargeSymbols}{bold}{OMX}{MnSymbolE}{b}{n}
\DeclareFontShape{OMX}{MnSymbolE}{m}{n}{
    <-6>  MnSymbolE5
   <6-7>  MnSymbolE6
   <7-8>  MnSymbolE7
   <8-9>  MnSymbolE8
   <9-10> MnSymbolE9
  <10-12> MnSymbolE10
  <12->   MnSymbolE12
}{}
\DeclareFontShape{OMX}{MnSymbolE}{b}{n}{
    <-6>  MnSymbolE-Bold5
   <6-7>  MnSymbolE-Bold6
   <7-8>  MnSymbolE-Bold7
   <8-9>  MnSymbolE-Bold8
   <9-10> MnSymbolE-Bold9
  <10-12> MnSymbolE-Bold10
  <12->   MnSymbolE-Bold12
}{}

\let\llangle\@undefined
\let\rrangle\@undefined
\DeclareMathDelimiter{\llangle}{\mathopen}%
                     {MnLargeSymbols}{'164}{MnLargeSymbols}{'164}
\DeclareMathDelimiter{\rrangle}{\mathclose}%
                     {MnLargeSymbols}{'171}{MnLargeSymbols}{'171}
\makeatother

\DeclareMathOperator\iden{iden}
\DeclareMathOperator\comp{comp}
\DeclareMathOperator\unit{unit}
\DeclareMathOperator\injl{injl}
\DeclareMathOperator\injr{injr}
\DeclareMathOperator\case{case}
\DeclareMathOperator\assertl{assertl}
\DeclareMathOperator\assertr{assertr}
\DeclareMathOperator\pair{pair}
\DeclareMathOperator\drop{drop}
\DeclareMathOperator\take{take}
\DeclareMathOperator\fail{fail}
\DeclareMathOperator\witness{witness}
\DeclareMathOperator\checkSig{checkSig}
\DeclareMathOperator\basicVerify{basicSigVerify}
\DeclareMathOperator\pubKey{pubKey}
\DeclareMathOperator\sigHash{sighash}
\DeclareMathOperator\makeSigHash{MakeSigHash}
\DeclareMathOperator\SigHashType{SigHashType}
\DeclareMathOperator\PubKey{PubKey}
\DeclareMathOperator\Signature{Signature}
\DeclareMathOperator\Transaction{Trans}
\DeclareMathOperator\flip{not}
\DeclareMathOperator\adder{half-adder}
\DeclareMathOperator\fulladder{full-adder}
\DeclareMathOperator\sha{sha-256-block}
\DeclareMathOperator\bitSize{bitSize}
\DeclareMathOperator\padl{padl}
\DeclareMathOperator\padr{padr}
\DeclareMathOperator\ecb{extraCellBnd}
\DeclareMathOperator\cb{cellBnd}
\DeclareMathOperator\tg{tag}
\DeclareMathOperator\compress{SHA256Block}
\DeclareMathOperator\tco{tco}
\DeclareMathOperator\on{on}
\DeclareMathOperator\off{off}
\DeclareMathOperator\monad{\mathfrak M}
\newcommand{\splat}{\langle{}\rangle}
\newcommand{\sigmaL}{\sigma^{\mathbf L}}
\newcommand{\sigmaR}{\sigma^{\mathbf R}}
\newcommand{\denote}[1]{\llbracket{#1}\rrbracket}
\newcommand{\denoteM}[1]{\llbracket{#1}\rrbracket_{\monad}}
\newcommand{\interp}[2]{\lceil{#1}\rceil_{#2}}
\newcommand{\repr}[2]{\lfloor{#1}\rfloor_{#2}}
\newcommand{\cells}[2]{\ulcorner{#1}\urcorner_{#2}}
\newcommand{\tuple}[2]{\langle{#1},{#2}\rangle}
\newcommand{\bm}[1]{\llangle{#1}\rrangle^{*}}
\newcommand{\tcon}[1]{\llangle{#1}\rrangle^{\tco}_{\on}}
\newcommand{\tcoff}[1]{\llangle{#1}\rrangle^{\tco}_{\off}}
\newcommand{\mr}[1]{\#({#1})}
\newcommand{\ecbtco}{\ecb^{\tco}}
\newcommand{\cbtco}{\cb^{\tco}}
\begin{document}

\title{Simplicity: A New Language for Blockchains}
\author{Russell O'Connor \\ {roconnor@blockstream.com}}
\date{2017-12-13}

\maketitle

\begin{abstract}
Simplicity is a typed, combinator-based, functional language without loops and recursion, designed to be used for crypto-currencies and blockchain applications.
It aims to improve upon existing crypto-currency languages, such as Bitcoin Script and Ethereum's EVM, while avoiding some of the problems they face.
Simplicity comes with formal denotational semantics defined in Coq, a popular, general purpose software proof assistant.
Simplicity also includes operational semantics that are defined with an abstract machine that we call the Bit Machine.
The Bit Machine is used as a tool for measuring the computational space and time resources needed to evaluate Simplicity programs.
Owing to its Turing incompleteness, Simplicity is amenable to static analysis that can be used to derive upper bounds on the computational resources needed, prior to execution.
While Turing incomplete, Simplicity can express any finitary function, which we believe is enough to build useful ``smart contracts'' for blockchain applications.
\end{abstract}

%\keywords{blockchain; bounded computation; crypto-currency; formal semantics; smart contracts}

\section{License}
    \begin{center}
      \begin{minipage}{\textwidth-10em}
        \doclicenseLongText%
        \textlangle\doclicenseURL\textrangle
      \end{minipage}
      \hfill
      \begin{minipage}{8em}
        \doclicenseImage%
      \end{minipage}
    \end{center}

\section{Introduction}

Blockchain and distributed ledger technologies define protocols that allow large, ad-hoc, groups of users to transfer value between themselves without needing to trust each other or any central authority.
Using public-key cryptography, users can sign transactions that transfer ownership of funds to other users.
To prevent transactions from conflicting with each other, for example when one user attempts to transfer the same funds to multiple different users at the same time, a consistent sequence of blocks of transactions is committed using a proof of work scheme.
This proof of work is created by participants called miners.
Each user verifies every block of transactions; among multiple sequences of valid blocks, the sequence with the most proof of work is declared to be authoritative.

Bitcoin~\cite{bitcoin} was the first protocol to use this blockchain technology to create a secure and permissionless crypto-currency.
It comes with a built-in programming language, called Bitcoin Script~\cite{script,satoshiScript}, which determines if a transaction is authorized to transfer funds.
Rather than sending funds to a specific party, users send funds to a specific Bitcoin Script program.
This program guards the funds and only allows them to be redeemed by a transaction input that causes the program to return successfully.
Every peer in the Bitcoin network must execute each of these programs with the input provided by the transaction data and all these programs must return successfully in order for the given transaction to be considered valid.

A typical program specifies a public key and simply requires that a valid digital signature of the transaction for that key be provided.
Only someone with knowledge of the corresponding private key can create such a signature.

Funds can be guarded by more complex programs.
Examples range from simpler programs that require signatures from multiple parties, permitting escrow services~\cite{goldfederescrow}, to more complex programs that allow for zero-knowledge contingent payments~\cite{zkcp}, allowing for trustless and atomic purchases of digital data.
This last example illustrates how Bitcoin Script can be used to build smart contracts, which allow parties to create conditional transactions that are enforced by the protocol itself.

Because of its Script language, Bitcoin is sometimes described as programmable money.

\subsection{Bitcoin Script}

Bitcoin Script is a stack-based language similar to Forth.
A program in Bitcoin Script is a sequence of operations for its stack machine.
Bitcoin Script has conditionals but no loops, thus all programs halt and the language is not Turing complete.

Originally, these programs were committed as part of the output field of a Bitcoin transaction.
The program from the output being redeemed is executed starting with the initial stack provided by the input redeeming it.
The program is successful if it completes without crashing and leaves a non-zero value on the top of stack.

Later, a new option called \textit{pay to script hash} (P2SH)~\cite{andresen2012-2} was added to Bitcoin.
Using this option, programs are committed by specifying a hash of the source code as part of the output field of a Bitcoin transaction.
The input field of a redeeming Bitcoin transaction then provides both the program and the initial stack for the program.
The transaction is valid if the hash of the provided program matches the hash specified in the output being redeemed, and if that program executes successfully with the provided input, as before.
The main effect here is moving the cost of large programs from the person sending funds to the person redeeming funds.

Bitcoin Script has some drawbacks.
Many operations were disabled by Bitcoin's creator, Satoshi Nakamoto~\cite{misc}.
This has left Bitcoin Script with a few arithmetic (multiplication was disabled), conditional, stack manipulation, hashing, and digital-signature verification operations.
In practice, almost all programs found in Bitcoin follow a small set of simple templates to perform digital signature verification.

However, Bitcoin Script also has some desirable properties.
All Bitcoin Script operations are pure functions of the machine state except for the signature-verification operations.
These signature-verification operations require a hash of some of the transaction data.
Together, this means the program's success or failure is purely a function of the transaction data.
Therefore, the person creating a transaction can know whether the transaction they have created is valid or not.

Bitcoin Script is also amenable to static analysis, which is another desirable property.
The digital-signature verification operations are the most expensive operations.
Prior to execution, Bitcoin counts the number of occurrences of these operations to compute an upper bound on the number of expensive calls that may occur.
Programs whose count exceeds a certain threshold are invalid.
This is a form of static analysis that ensures the amount of computation that will be done isn't excessive.
Moreover, before committing a program to an output, the creator of the program can ensure that the amount of computation that will be done will not be excessive.

\subsection{Ethereum and the EVM}

Ethereum~\cite{wood2014} is another crypto-currency with programmable transactions.
It defines a language called EVM for its transactions.
While more expressive and flexible than Bitcoin Script, the design of Ethereum and the EVM has several issues.

The EVM is a Turing-complete programming language with a stack, random access memory, and persistent storage.
To prevent infinite loops, execution is limited by a counter called gas, which is paid for in Ethereum's unit of account, Ether, to the miner of the block containing the transaction.
Static analysis isn't practical with a Turing-complete language.
Therefore, when a program runs out of gas, the transaction is nullified but the gas is still paid to the miner to ensure they are compensated for their computation efforts.

Because of direct access to persistent storage, EVM programs are a function of both the transaction and the state of the blockchain at the point the transaction is inserted into the blockchain.
This means users cannot necessarily know what the result of their transaction will be when they create it, nor can they necessarily know how much gas will be consumed.
Users must provide enough gas to cover the worst-case use scenario and there is no general purpose algorithm that can compute that bound.
In particular, there is no practical general purpose way to tell if any given program will run out of gas in some context either at redemption time or creation time.
There have been some efforts to perform static analysis on a subset of the EVM language~\cite{Bhargavan:2016}, but these tools are limited (e.g. they do not support programs with loops). 

Unlike Bitcoin, there are many instances of ad-hoc, one-off, special purpose programs in Ethereum.
These programs are usually written in a language called Solidity~\cite{solidity} and compiled to the EVM.
These ad-hoc programs are regularly broken owing to the complex semantics of both Solidity and the EVM; the most famous of these failures were the DAO~\cite{DAO-hack} and Parity's multiple signature validation program~\cite{parityWallet}.

\subsection{A New Language}

In this paper, we propose a new language that maintains or enhances the desirable properties that Bitcoin Script has while adding expressiveness.
Our design goals are:
\begin{itemize}
\item Create an expressive language that provides users with the tools needed to build novel programs and smart contracts.
\item Enable static analysis that provides useful upper bounds on the amount of computation required.
\item Minimize bandwidth and storage requirements and enhance privacy by removing unused code at redemption time.
\item Maintain Bitcoin's design of self-contained transactions whereby programs do not have access to any information outside the transaction.
\item Provide formal semantics that facilitate easy reasoning about programs using existing off-the-shelf proof-assistant software.
\end{itemize}

A few of these goals deserve additional explanation:

Static analysis allows the protocol to place limits on the amount of computation a transaction can have, so that nodes running the protocol are not overly burdened.
Furthermore, the static analysis can provide program creators with a general purpose tool for verifying that the programs they build will always fit within these limits.
Additionally, it is easy for the other participants in a contract to check the bounds on the smart contract's programs themselves.

Self-contained transactions ensure that execution does not depend on the global state of the blockchain.
Once a transaction's programs have been validated, that fact can be cached.
This is particularly useful when transactions are being passed around the network before inclusion in the blockchain;
once included in the blockchain, the programs do not need to be executed again.

Finally, formal semantics that work with proof-assistant software provide the opportunity for contract developers to reason about their programs to rule out logical errors and to help avoid scenarios like the DAO and Parity's multi-signature program failure.

Like Bitcoin Script and the EVM, our language is designed as low-level language for executing smart contracts, not as a language for coding in directly.
As such, we expect it to be a target for other, higher-level, languages to be compiled to.

We call our new language \textit{Simplicity}.

\section{Core Simplicity}

Simplicity is a typed combinator language.
Each well-typed Simplicity expression is associated with two types, an input type and an output type.
Every expression denotes a function from its input type to its output type.

\subsection{Types in Simplicity}

\begin{figure*}[t]
\[
\begin{array}{cc}
\inferrule {{}} {\iden : A \vdash A}
&
\inferrule {s : A \vdash B \\ t : B \vdash C} {\comp s\, t : A \vdash C}
\\
\multicolumn{2}{c}{\inferrule {{}} {\unit : A \vdash {\mathbb 1}}}
\\\\
\inferrule {t : A \vdash B} {\injl t : A \vdash B + C}
&
\inferrule {t : A \vdash C} {\injr t : A \vdash B + C}
\\\\
\inferrule {s : A \times C \vdash D \\ t : B \times C \vdash D} {\case s\, t : (A + B) \times C \vdash D}
&
\inferrule {s : A \vdash B \\ t : A \vdash C} {\pair s\, t : A \vdash B \times C}
\\\\
\inferrule {t : A \vdash C} {\take t : A \times B \vdash C}
&
\inferrule {t : B \vdash C} {\drop t : A \times B \vdash C}
\end{array}
\]
\caption{Typing rules for the terms of core Simplicity.}
\label{TyCore}
\end{figure*}

Types in Simplicity come in three flavors.

\begin{itemize}
 \item The unit type, written as ${\mathbb 1}$, is the type with one element.
 \item A sum type, written as $A + B$, contains the tagged union of values from either the left type $A$ or the right type $B$.
 \item A product type, written as $A \times B$, contains pairs of elements with the first one from the type $A$ and the second one from the type $B$.
\end{itemize}

There are no recursive types in Simplicity.
Every type in Simplicity only contains a finite number of values;
the number of possible values a type contains can be computed by interpreting the type as an arithmetic expression.

\subsection{Terms in Simplicity}

The term language for Simplicity is based on Gentzen's sequent calculus~\cite{gentzen}.
We write a type-annotated Simplicity expression as $t : A \vdash B$ where $t$ a Simplicity expression, $A$ is the expression's input type, and $B$ is the expression's output type.

The core of Simplicity consists of nine combinators for building expressions.
They capture the fundamental operations for the three flavors of types in Simplicity.
The typing rules for these nine combinators is given in Figure~\ref{TyCore}.

We can classify the combinators based on the flavor of types they support.
\begin{itemize}
\item The $\unit$ term returns the singular value of the unit type and ignores its argument.
\item The $\injl$ and $\injr$ combinators create tagged values, while the $\case$ combinator, Simplicity's branching operation, evaluates one of its two sub-expressions based on the tag of the first component of its input.
\item The $\pair$ combinator creates pairs, while the $\take$ and $\drop$ combinators access first and second components of a pair respectively.
\item The $\iden$ and $\comp$ combinators are not specific to any flavor of type. The $\iden$ term represents the identity function for any type and the $\comp$ combinator provides function composition.
\end{itemize}
Simplicity expressions form an abstract syntax tree.
The leaves of this tree are either $\iden$ or $\unit$ terms.
The nodes are one of the other seven combinators.
Each node has one or two children depending on which combinator the node represents.

Precise semantics are given in the next section.
Extensions to this core set of combinators is given in Section~\ref{extSimplicity}.

\subsection{Denotational Semantics of Simplicity}

Before giving the semantics of Simplicity terms, we need notations for the values of Simplicity types.

\begin{itemize}
\item We write $\splat : {\mathbb 1}$ for the singular value of the unit type.
\item We write $\sigmaL(a) : A + B$ when $a : A$ for left-tagged values of the sum type.
\item We write $\sigmaR(b) : A + B$ when $b : B$ for right-tagged values of the sum type.
\item We write $\tuple{a}{b} : A \times B$ when $a : A$ and $b : B$ for values of the pair type.
\end{itemize}

We emphasize that these values are not directly representable in Simplicity because Simplicity expressions can only denote functions;
we use these values only to define the functional semantics of Simplicity.

Simplicity's denotational semantics are recursively defined.
For an expression $t : A \vdash B$, we define its semantics $\denote{t} : A \rightarrow B$ as follows.

\begin{align*}
\denote{\iden}(a) &\coloneqq a\\
\denote{\comp s\, t}(a) &\coloneqq \denote{t}(\denote{s}(a))\\
\denote{\unit}(a) &\coloneqq \splat\\
\denote{\injl t}(a) &\coloneqq \sigmaL(\denote{t}(a))\\
\denote{\injr t}(a) &\coloneqq \sigmaR(\denote{t}(a))\\
\denote{\case s\, t}\tuple{\sigmaL(a)}{c} &\coloneqq \denote{s}\tuple{a}{c}\\
\denote{\case s\, t}\tuple{\sigmaR(b)}{c} &\coloneqq \denote{t}\tuple{b}{c}\\
\denote{\pair s\, t}(a) &\coloneqq \tuple{\denote{s}(a)}{\denote{t}(a)}\\
\denote{\take t}\tuple{a}{b} &\coloneqq \denote{t}(a)\\
\denote{\drop t}\tuple{a}{b} &\coloneqq \denote{t}(b)\\
\end{align*}

We have formalized the language and semantics of core Simplicity in the Coq proof assistant~\cite{coq}.
The formal semantics in Coq are the official semantics of Simplicity, and for core Simplicity, it is short enough that it fits in Appendix~\ref{FormalDef} of this paper.
Having semantics in a general purpose proof assistant allows us to formally reason both about programs written in Simplicity and about algorithms that analyze Simplicity programs.

\subsection{Completeness}

Simplicity cannot express general computation.
It can only express finitary functions, because each Simplicity type contains only finitely many values.
However, within this domain, Simplicity's set of combinators is complete:
any function between Simplicity's types can be expressed.

\begin{theorem}[Finitary Completeness]
Let $A$ and $B$ be Simplicity types.
Let $f : A \rightarrow B$ be any function between these types.
Then there exists some core Simplicity expression $t : A \vdash B$ such that $\denote{t} = f$.
\end{theorem}
We have verified this theorem with Coq.

This theorem holds because functions between finite types can be described, in principle, by a lookup table that maps inputs to outputs.
In principle, such a lookup table can be encoded as a Simplicity expression that does repeated case analysis on its input and returns a fixed-output value for each possible case.

However, using this theorem to construct Simplicity expressions is not practical.
For example, a lookup table for the compression function for SHA-256~\cite{sha}, which maps 768 bits to 256 bits, would be astronomical in size, requiring $2^{776}$ bits.
To create practical programs in Simplicity, we need to take advantage of structured computation in order to succinctly express functions.

\subsection{Example Programs}

The combinators of core Simplicity may seem paltry, so in this section we illustrate how we can construct programs in Simplicity.
We begin by defining a type for a bit, ${\mathbb 2}$, as the sum of two unit types.

\[{\mathbb 2} \coloneqq {\mathbb 1} + {\mathbb 1}\]

We choose an interpretation of bits as numbers where we define the left-tagged value as denoting zero and the right tagged value as denoting one.
\begin{align*}
  \interp{\sigmaL{\splat}}{\mathbb 2} &\coloneqq 0\\
  \interp{\sigmaR{\splat}}{\mathbb 2} &\coloneqq 1
\end{align*}

We can write Simplicity programs to manipulate bits.
For example, we can define the $\flip$ function to flip a bit.
\begin{align*}
 \flip &: {\mathbb 2} \vdash {\mathbb 2}\\
 \flip &\coloneqq \comp\,(\pair \iden \unit)\,(\case\,(\injr \unit)\,(\injl \unit))
\end{align*}

By recursively taking products, we can define types for multi-bit words.

\begin{align*}
{\mathbb 2}^1 &\coloneqq {\mathbb 2}\\
{\mathbb 2}^{2n} &\coloneqq {\mathbb 2}^n \times {\mathbb 2}^n 
\end{align*}

For example, the ${\mathbb 2}^{32}$ type can represent 32-bit words. 

We recursively define the interpretation of pairs of words as numbers, choosing to use big-endian order.
\[
  \interp{\tuple{a}{b}}{{\mathbb 2}^{2n}} \coloneqq \interp{a}{{\mathbb 2}^n} \cdot 2^n + \interp{b}{{\mathbb 2}^n}
\]

We can write a half-adder of two bits in Simplicity.
\begin{align*}
 \adder &: {\mathbb 2} \times {\mathbb 2} \vdash {\mathbb 2}^2\\
 \adder &\coloneqq\case\,\begin{aligned}[t]
        &(\drop\,(\pair\,(\injl \unit)\, \iden))\\
	&(\drop\,(\pair \iden\, \flip))\\
        \end{aligned}
\end{align*}

We can prove the $\adder$ expression correct.
\begin{theorem}[Half Adder Correct]
Let $a, b : {\mathbb 2}$ be two bits.
Then 
\[
\interp{\denote{\adder}\tuple{a}{b}}{{\mathbb 2}^2} = \interp{a}{\mathbb 2} + \interp{b}{\mathbb 2}
\]
\end{theorem}

We have proven this theorem in Coq.
This particular theorem can be proven by exhaustive case analysis (there are only four cases) and equational reasoning using Simplicity's denotational semantics.

We continue and define a full adder by combining two half adders.
From there we can build ripple-carry adders for larger word sizes by combining full adders from smaller word sizes.
The interested reader can find the Simplicity expressions for these full adders in Appendix~\ref{FullAdder}.

Continuing, we can build multipliers and other bit-manipulation functions.
These can be combined to implement more sophisticated functions.

We have written the SHA-256 block compression function in Simplicity.
Using 256-bit arithmetic, we have also constructed expressions for elliptic curve operations over the Secp256k1 curve~\cite{sec2} that Bitcoin uses, and
we have built a form of ECDSA signature validation~\cite{ecdsa} in Simplicity.

% Prelude Sequent Sha256> termSize sha256Block 
% 3274442
% Prelude Sequent Sha256> Data.Map.size (termHashSet sha256Block)
% 865
% Prelude Sequent Serialize Sha256> Data.Map.size (wideFootprintMap sha256Block)
% 1130
To gain an understanding of the size of Simplicity expressions, let us examine our implementation of the SHA-256 block compression function, $\sha : {\mathbb 2}^{256} \times {\mathbb 2}^{512} \vdash {\mathbb 2}^{256}$.
Our Simplicity expression consists of 3\,274\,442 combinators.
However, this counts the total number of nodes in the abstract syntax tree that makes up the expression.
Several sub-expressions of this expression are duplicated and can be shared.
Imagine taking the abstract syntax tree and sharing identical sub-expressions to create a directed acyclic graph (DAG) representing the same expression.
Counting this way, the same expression contains only 1\,130 unique typed sub-expressions, which is the number of nodes that would occur in the DAG representing the expression.

Taking advantage of shared sub-expressions is critical because it makes the representation of typical expressions exponentially smaller in size.
We choose to leave sub-expression sharing implicit in Simplicity's term language.
This ensures that Simplicity's semantics are not affected by how sub-expressions are shared.
However, we do transmit and store Simplicity expressions in a DAG format.
This DAG representation of expressions is also important when we consider static analysis in Section~\ref{staticAnalysis}.

Using our formal semantics of Simplicity in Coq, we have proven our implementation of SHA-256 is correct.

\begin{theorem}[SHA-256 Correct]
Let $a : {\mathbb 2}^{256}$ and $b : {\mathbb 2}^{512}$.
Let\linebreak{}$\sha : {\mathbb 2}^{256} \times {\mathbb 2}^{512} \vdash {\mathbb 2}^{256}$ be our Simplicity implementation of the SHA-256 block compression function.
Let $\compress : {\mathbb 2}^{256} \times {\mathbb 2}^{512} \rightarrow {\mathbb 2}^{256}$ be the SHA-256 block compression function.
Then 
\[
\denote{\sha}{\tuple{a}{b}} = \compress\tuple{a}{b}
\]
\end{theorem}

Our reference implementation of the SHA-256 compression function in Coq\footnote{The type of the reference implementation for the SHA-256 compression function in Coq actually uses lists of 32-bit words.
To simplify the presentation here, we have omitted the translation between representations of the compression function's inputs and outputs.}
is taken from the Verified Software Toolchain (VST) project~\cite{appel} where they use it as their reference implementation for proving OpenSSL's C implementation of SHA-256 correct~\cite{Appel:2015}.
If we combine our Coq proof with the VST project's Coq proof, we would get a formal proof that our Simplicity implementation of SHA-256 matches OpenSSL's C implementation of SHA-256.

\section{The Bit Machine}

One of the design goals for Simplicity is to be able to compute a bound on the computation costs of evaluating a Simplicity program.
While the denotational semantics of Simplicity are adequate for determining the functional behavior of Simplicity programs, we need operational semantics to provide a measure of computational resources used.
To do this, we create an abstract machine, called the Bit Machine, tailored for evaluating Simplicity programs.

The Bit Machine is an abstract imperative  machine whose state consists of two non-empty stacks of data frames.
One stack holds read-only frames, and the other stack holds write-only frames.
A data frame consists of an array of cells and a cursor pointing either at one of the cells or pointing just beyond the end of the array.
Each cell contains one of three values: a zero bit, a one bit, or an undefined value.
The Bit Machine has a set of instructions that manipulate the two stacks and their data frames.

Notionally, we will write a frame like in the following example.

\[
\mathtt{[01\underline{0}?1?1]}
\]

This frame contains 7 cells.
The $\mathtt{?}$ character denotes a cell with an undefined value.
The $\mathtt{0}$ and $\mathtt{1}$ characters denotes cells with the corresponding bit value.
The frame's cursor is pointing to the underscored cell.

When a frame's cursor is pointing past the end of the array, we will write an underscore after the array's cells like the following.

\[
\mathtt{[010?1?1\underline{]}}
\]

Table~\ref{bmstate} illustrates an example of a possible state of the Bit Machine.
This example contains a read-frame stack that has four data frames and a write-frame stack that has two data frames.
The top frame of the read-frame stack is called the \textit{active read frame}, and the top frame of the write-frame stack is called the \textit{active write frame}.

\begin{table}[h]\centering
\begin{tabular}{@{}ll@{}}\toprule
read frame stack & write frame stack\\\midrule
$\mathtt{[100\underline{1}1??110101000]}$&$\mathtt{[11??1101\underline{]}}$\\
$\mathtt{[\underline{0}000]}$&$\mathtt{[111\underline{?}?]}$\\
$\mathtt{[\underline{]}}$\\
$\mathtt{[\underline{1}0]}$\\
\bottomrule
\end{tabular}
\caption{Example state for the Bit Machine}\label{bmstate}
\end{table}

The Bit Machine has ten instructions to manipulate its state, which we describe below.

\begin{itemize}
\item $\mathbf{nop}$: This instruction doesn't change the state of the machine.
\item $\mathbf{write}(b)$:
This instruction writes bit value $b$ to the cell under the active write frame's cursor and advances the cursor by one cell.
\begin{itemize}
 \item The value $b$ must be either $\mathtt{0}$ or $\mathtt{1}$ or else the machine crashes.
 \item The active write frame's cursor must not be beyond the end of the array or else the machine crashes.
 \item The cell written to must have an undefined value before writing to it or else the machine crashes.
 \end{itemize}
\item $\mathbf{copy}(n)$:
This instruction copies $n$ cells from the active read frame, starting at its cursor, to the active write frame, starting at its cursor, and advances the active write frame's cursor by $n$ cells.
\begin{itemize}
 \item There must be at least $n$ cells between the active read frame's cursor and the end of its array or else the machine crashes.
 \item There must be at least $n$ cells between the active write frame's cursor and the end of its array and they all must have an undefined value before writing to them or else the machine crashes.
 \item Note that undefined values can be copied.
 \end{itemize}
\item $\mathbf{skip}(n)$:
This instruction advances the active write frame's cursor by $n$ cells.
\begin{itemize}
 \item This instruction is allowed to advance the cursor to the point just past the end of the array; however if $n$ is large enough that it would advance it past this point, the machine crashes instead.
 \end{itemize}
\item $\mathbf{fwd}(n)$:
This instruction moves the active read frame's cursor forward by $n$ cells.
\begin{itemize}
 \item This instruction is allowed to advance the cursor to the point just past the end of the array; however if $n$ is large enough that it would advance it past this point, the machine crashes instead.
 \end{itemize}
\item $\mathbf{bwd}(n)$:
This instruction moves the active read frame's cursor backwards by $n$ cells.
\begin{itemize}
 \item If this instruction would move the cursor before the beginning of the array, the machine crashes instead.
 \end{itemize}
\item $\mathbf{newFrame}(n)$:
This instruction allocates a new frame with $n$ cells and pushes it onto the write-frame stack, making it the new active write frame.
\begin{itemize}
 \item The cursor for this new frame starts at the beginning of the array.
 \item The cells of the new frame are initialized with undefined values.
 \end{itemize}
\item $\mathbf{moveFrame}$:
This instruction pops a frame off the write-frame stack and pushes it onto the read-frame stack.
\begin{itemize}
 \item The moved frame has its cursor reinitialized to the beginning of the array.
 \item If this would leave the machine with an empty write frame stack, the machine crashes instead.
 \end{itemize}
\item $\mathbf{dropFrame}$:
This instruction pops a frame off the read-frame stack and deallocates it.
\begin{itemize}
 \item If this would leave the machine with an empty read-frame stack, the machine crashes instead.
 \end{itemize}
\item $\mathbf{read}$:
This instruction doesn't modify the state of the machine.
Instead it returns the value of the bit under the active read frame's cursor to the machine's operator.
\begin{itemize}
 \item if the value under the active read frame's cursor is undefined, or if it is past the end of its array, the machine crashes instead.
\end{itemize}
\end{itemize}

The Bit Machine is deliberately designed to crash at anything resembling undefined behavior.

\subsection{Bit Machine Values}

We can use the Bit Machine to evaluate Simplicity programs on their inputs.
First, we define how many cells are needed to hold the values of a given type.

\begin{align*}
  \bitSize({\mathbb 1}) &\coloneqq 0\\
  \bitSize(A + B) &\coloneqq 1 + \max(\bitSize(A), \bitSize(B))\\
  \bitSize(A \times B) &\coloneqq \bitSize(A) + \bitSize(B)
\end{align*}

The unit type doesn't need any bits to represent its one value.
A value of sum type needs one bit for its tag and then enough bits to represent either of its left or right values.
A value of product type needs space to hold both its values.

We precisely define a representation of values for Simplicity types as array of cells below:

\begin{align*}
  \cells{\splat}{\mathbb 1} &\coloneqq \mathtt{[]}\\
  \cells{\sigmaL{(a)}}{A + B} &\coloneqq \mathtt{[0]}\cdot\mathtt{[?]}^{\padl(A,B)}\cdot\cells{a}{A}\\
  \cells{\sigmaR{(b)}}{A + B} &\coloneqq \mathtt{[1]}\cdot\mathtt{[?]}^{\padr(A,B)}\cdot\cells{b}{B}\\
  \cells{\tuple{a}{b}}{A \times B} &\coloneqq \cells{a}{A}\cdot\cells{b}{B}
\end{align*}

For the notation above, $\cdot$ denotes the concatenation of arrays of cells and $\mathtt{[}b\mathtt{]}^n$ denotes an array of $n$ cells all containing $b$.

The padding for left and right values fills the array with sufficient undefined values so that
the array ends up with the required length.

\begin{align*}
  \padl(A,B) &\coloneqq \max(\bitSize(A),\bitSize(B)) - \bitSize(A)\\
  \padr(A,B) &\coloneqq \max(\bitSize(A),\bitSize(B)) - \bitSize(B)\\
\end{align*}

Below are some examples of values represented as arrays of cells for the Bit Machine.
We define the inverse of $\interp{\cdot}{{\mathbb 2}^n}$ to be $\repr{\cdot}{{\mathbb 2}^n}$ that maps numbers to the values that represent them.
\begin{align*}
  \cells{\sigmaL(\repr{3}{{\mathbb 2}^2})}{{\mathbb 2}^2 + {\mathbb 2}} &= \mathtt{[011]}\\
  \cells{\sigmaR(\repr{0}{{\mathbb 2}})}{{\mathbb 2}^2 + {\mathbb 2}} &= \mathtt{[1?0]}
\end{align*}

The array of cells contains a list of the tags from the sum types and that is about it.
Notice that values of a multi-bit word type, ${\mathbb 2}^n$, have exactly $n$ cells and the representation is identical to its big-endian binary representation.

\subsection{Operational Semantics}

The operational semantics for Simplicity are given by recursively translating a Simplicity expression into a sequence of instructions for the Bit Machine.
Figure~\ref{bmdef} defines $\bm{t : A \vdash B}$, which results in a sequence of instructions for the Bit Machine that evaluates the function $\denote{t}$.

\begin{figure*}[t]
\begin{align*}
 \bm{\iden : A \vdash A} &\coloneqq \mathbf{copy}(\bitSize(A))\\
 \bm{\comp s\,t : A \vdash C} &\coloneqq
   \begin{aligned}[t]
   &\mathbf{newFrame}(\bitSize(B));\bm{s : A \vdash B};\\
   &\mathbf{moveFrame};\bm{t : B \vdash C};\\
   &\mathbf{dropFrame}
   \end{aligned}\\
 \bm{\unit : A \vdash {\mathbb 1}} &\coloneqq \mathbf{nop}\\
 \bm{\injl t : A \vdash B + C} &\coloneqq \mathbf{write}(\mathtt{0});\mathbf{skip}(\padl(B,C));\bm{t : A \vdash B}\\
 \bm{\injr t : A \vdash B + C} &\coloneqq \mathbf{write}(\mathtt{1});\mathbf{skip}(\padr(B,C));\bm{t : A \vdash C}\\
 \bm{\case s\,t : (A + B) \times C \vdash D} &\coloneqq \begin{aligned}[t]
   &\text{match }\mathbf{read}\text{ with}\\
   &\begin{cases}
   \texttt{0} \rightarrow \begin{aligned}[t]
                          &\mathbf{fwd}(1 + \padl(A,B));\\
                          &\bm{s : A \times C \vdash D};\\
                          &\mathbf{bwd}(1 + \padl(A,B))
                          \end{aligned}\\
   \texttt{1} \rightarrow \begin{aligned}[t]
                          &\mathbf{fwd}(1 + \padr(A,B));\\
                          &\bm{t : B \times C \vdash D};\\
                          &\mathbf{bwd}(1 + \padr(A,B))
                          \end{aligned}
   \end{cases}
   \end{aligned}\\
 \bm{\pair s\,t : A \vdash B \times C} &\coloneqq \bm{s : A \vdash B};\bm{t : A \vdash C}\\
 \bm{\take t : A \times B \vdash C} &\coloneqq \bm{t : A \vdash C}\\
 \bm{\drop t : A \times B \vdash C} &\coloneqq \begin{aligned}[t]
                                               &\mathbf{fwd}(\bitSize(A));\bm{t : B \vdash C};\\
                                               &\mathbf{bwd}(\bitSize(A))
                                               \end{aligned}\\
\end{align*}
\caption{Operational Semantics for Simplicity using the Bit Machine.}\label{bmdef}
\end{figure*}

We have formalized the Bit Machine in Coq and Figure~\ref{bmdef}'s translation of Simplicity expressions into Bit Machine instructions.
We have verified, with Coq, that our Bit Machine implementation of Simplicity respects Simplicity's denotational semantics.

\begin{theorem}[Correctness of Operational Semantics]\label{cos}
Let $t : A \vdash B$ be any Simplicity expression.
Let $v : A$ be any value of type $A$.
Initialize a Bit Machine with 
\begin{itemize}
\item the value $\cells{v}{A}$ as the only frame on the read-frame stack with its cursor at the beginning of the frame, and
\item the value $\mathtt{[?]}^{\bitSize(B)}$ as the only frame on the write-frame stack with its cursor at the beginning of the frame.
\end{itemize}
After executing the instructions $\bm{t : A \vdash B}$, the final state of the Bit Machine has
\begin{itemize}
\item the value $\cells{v}{A}$ as the only frame on the read-frame stack, with its cursor at the beginning of the frame, and
\item the value $\cells{\denote{t}(v)}{B}$ as the only frame on the write-frame stack, with its cursor at the end of the frame.
\end{itemize}
In particular, the Bit Machine never crashes during this execution.
\end{theorem}

The fact that the Bit Machine never crashes means that during execution it never reads from an undefined cell, nor does it ever write to a defined cell.
This means that, if one implements the Bit Machine on a real computer, one can use an ordinary array of bits to represent the array of cells and cells that are supposed to hold undefined values can be safely set to any bit value.
As long as this implementation only executes instructions translated from Simplicity and started from a proper initial state, the resulting computation will be the same as if one had used an explicit representation for undefined values.

\subsection{Static Analysis}\label{staticAnalysis}
Using the Bit Machine, we can measure computational resources needed by a Simplicity program.
For instance we can:
\begin{itemize}
\item count the number of instructions executed by the Bit Machine.
\item count the number of cells copied by the Bit Machine.
\item count the maximum number of cells in both stacks at any point during execution.
\item count the maximum number of frames in both stacks at any point during execution.
\end{itemize}

The first two are measurements of time used by the Bit Machine, and the last two are measurements of space used by the Bit Machine.

Using simple static analysis, we can quickly compute an upper bound on these sorts of resource costs.
In this paper, we will focus on the example of counting the maximum number of cells in both stacks at any point during execution.
Figure~\ref{cbdef} defines a function $\cb$ that computes an upper bound on the maximum number of cells that are needed during execution.

\begin{figure*}[t]
\begin{align*}
 \ecb(\iden : A \vdash A) \coloneqq&\,0\\
 \ecb(\comp s\,t : A \vdash C) \coloneqq& \bitSize(B) +\\
 &\begin{aligned}[t]
 \max(\ecb(s : A \vdash B),\\
 \ecb(t : B \vdash C))
 \end{aligned}\\
 \ecb(\unit : A \vdash {\mathbb 1)} \coloneqq&\,0\\
 \ecb(\injl t : A \vdash B + C) \coloneqq& \ecb(t : A \vdash B)\\
 \ecb(\injr t : A \vdash B + C) \coloneqq& \ecb(t : A \vdash C)\\
 \ecb(\case s\,t : (A + B) \times C \vdash D) \coloneqq& \begin{aligned}[t]
 \max(\ecb(s : A \times C \vdash D),\\
 \ecb(t : B \times C \vdash D))
 \end{aligned}\\
 \ecb(\pair s\,t : A \vdash B \times C) \coloneqq& \begin{aligned}[t]
 \max(\ecb(s : A \vdash B),\\
 \ecb(t : A \vdash C))
 \end{aligned}\\
 \ecb(\take t : A \times B \vdash C) \coloneqq& \ecb(t : A \vdash C)\\
 \ecb(\drop t : A \times B \vdash C) \coloneqq& \ecb(t : B \vdash C)\\
\end{align*}
\[
 \cb(t : A \vdash B) \coloneqq \bitSize(A) + \bitSize(B) + \ecb(t : A \vdash B)
\]
\caption{Definition of $\cb$.}\label{cbdef}
\end{figure*}

\begin{theorem}[Static Analysis of Cell Usage]
Let $t : A \vdash B$ be any Simplicity expression.
Let $v : A$ be any value of type $A$.
Initialize a Bit Machine as specified in Theorem~\ref{cos}.
The maximum number of cells in both stacks of the Bit Machine at any point during the execution of $\bm{t : A \vdash B}$ from this initial state never exceeds $\cb(t : A \vdash B)$.
\end{theorem}

We have formalized the above static analysis and proven the above theorem in Coq.

These kinds of static analyses are simple recursive functions of Simplicity expressions, and the intermediate results for sub-expressions can be shared.
By using a DAG for the in-memory representation of Simplicity expressions, we can transparently cache these intermediate results.
This means the time needed to compute static analysis is proportional to the size of the DAG representing the Simplicity expression, as opposed to the time needed for dynamic analysis such as evaluation, which may take time proportional to the size of the tree representing the Simplicity expression.

The Bit Machine is an abstract machine, so we can think of these sorts of static analyses as bounding abstract resource costs.
As long as the abstract resource costs are limited, then the resource costs of any implementation of Simplicity will also be limited.
These sorts of precise static analyses of resource costs are more sophisticated than what is currently available for Bitcoin Script.

Notice that the definition of $\cb$ does not directly reference the Bit Machine, so
limits on the bounds computed by these static analyses can be imposed on protocols that use Simplicity without necessarily requiring that applications use the Bit Machine for evaluation.
While we do use an implementation of the Bit Machine in our prototype Simplicity evaluator written in C, others are free to explore other models of evaluation.
For example, the Bit Machine copies data to implement the $\iden$ combinator, but another model of evaluation might create shared references to data instead.

\subsubsection{Tail Composition Optimization}

Tail call optimization is an optimization used in many languages where a procedure's stack frame is freed prior to a call to another procedure when that call is the last command of the procedure.
The $\comp$ combinator in Simplicity behaves much like a procedure call, and we can implement an analogous optimization for the translation of Simplicity to the Bit Machine.
Interested readers can find this tail composition optimized translation, $\tcoff{t : A \vdash B}$, defined in Appendix~\ref{app:tco}, along with a static analysis of its memory use.

\subsection{Jets}
Evaluation of a Simplicity expression with the Bit Machine recursively traverses the expression.
Before evaluating some sub-expression $t : A \vdash B$, the Bit Machine is always in a state where the active read frame is of the form
\[
  r_1 \cdot \cells{v}{A} \cdot r_2
\]
for some value $v : A$ and some arrays $r_1$ and $r_2$, and where the cursor is placed at the beginning of the $\cells{v}{A}$ array slice.

Furthermore the active write frame is of the form
\[
  w_1 \cdot \mathtt{[?]}^{\bitSize(B)} \cdot w_2
\]
for some arrays $w_1$ and $w_2$, and where the cursor is placed at the beginning of the $\mathtt{[?]}^{\bitSize(B)}$ array slice.

After the evaluation of the sub-expression $t : A \vdash B$, the active write frame is of the form
\[
  w_1 \cdot \cells{\denote{t}(v)}{B} \cdot w_2
\]
and where the cursor is placed at the beginning of the $w_2$ array slice.

Taking an idea found in Urbit~\cite{urbit}, we notice that if we recognize a familiar sub-expression $t : A \vdash B$, the interpreter may bypass the BitMachine's execution of $\bm{t : A \vdash B}$ and instead directly compute and write $\cells{\denote{t}(v)}{B}$ to the active write frame.
Following Urbit, we call such a familiar expression and the code that replaces it a \textit{jet}.

Jets are essential for making Simplicity a practical language.
For our Blockchain application we expect to have jets for at least the following expressions:
\begin{itemize}
\item arithmetic operations (addition, subtraction, and multiplication) from \mbox{8-bit} to 256-bit word size, 
\item comparison operations (less than, less than or equal to, equality) from \mbox{8-bit} to 256-bit word size, 
\item logical bitwise operation for 8-bit to 256-bit word sizes,
\item constant functions for every possible 8-bit word,
\item compression functions from hash functions such as SHA-256's compression function,
\item elliptic curve point operations for the Secp256k1 curve~\cite{sec2}, and
\item digital signature validation for ECDSA~\cite{ecdsa} or Schnorr~\cite{schnorr1989} signatures.
\end{itemize}

We take advantage of the fact that the representation of the values used for arithmetic expressions in the Bit Machine match the binary memory format for real hardware.
This lets us write these jets by directly reading from and writing to data frames.

Jets have several nice properties:
\begin{itemize}
\item Jets provide a formal specification of their behavior.
The implementation of a jet must produce output identical to the output that would be produced by the Simplicity expression being replaced.
There is no possibility for an ambiguous interpretation of what a jet computes.
\item Jets cannot accidentally introduce new behavior or new side effects because they can only replicate the behavior of Simplicity expressions.
To add new behavior to Simplicity we must explicitly extend Simplicity (see Section~\ref{extSimplicity}).
\item Jets are transparent when it comes to reasoning about Simplicity expressions.
Jets are logically equal to the code they replace.
Therefore, when proving properties of one's Simplicity code, jets can safely be ignored.
\end{itemize}
Naturally, we expect jetted expressions to have properties already proven and available; this will aid reasoning about Simplicity programs that make use of jets.

Because jets are transparent, the static analyses of resource costs are not affected by their existence.
To encourage the use of jets, we anticipate discounts to be applied to the resource costs of programs that use jets based on the estimated savings of using jets.

When a suitably rich set of jets is available, we expect the bulk of the computation specified by a Simplicity program to be made up of these jets, with only a few combinators used to combine the various jets.
This should bring the computational requirements needed for Simplicity programs in line with existing blockchain languages.
In light of this, one could consider Simplicity to be a family of languages, where each language is defined by a set of jets that provide computational elements tailored for its particular application.

\section{Integration with Blockchains}\label{extSimplicity}
Core Simplicity is language that does pure computation.
In order to interact with blockchains we extend Simplicity with new combinators.

\subsection{Transaction Digests}
The main operation used in Bitcoin Script is \textsf{OP\_CHECKSIG} which, given a public key, a digital signature, and a SigHash type~\cite{sighash},
generates a digest of the transaction in accordance with the SigHash type and validates that the digital signature for the digest is correct for the given public key.
This operation allows users to create programs that require their signatures for their transactions to be authorized.

In core Simplicity we can implement, and jet, the digital signature validation algorithm.
However, it is not possible to generate the transaction digest because core Simplicity doesn't have access to the transaction data.
To remedy this, we add a new primitive combinator to Simplicity
\[
\inferrule {} {\sigHash : \SigHashType \vdash {\mathbb 2}^{256}}
\]
where $\SigHashType \coloneqq ({\mathbb 1} + {\mathbb 2}) \times {\mathbb 2}$ is a Simplicity type suitable for encoding all possible SigHash types.

This combinator returns the digest of the transaction for the given SigHash type.
Together with the Simplicity implementation of digital signature verification, we can implement the equivalent of Bitcoin's \textsf{OP\_CHECKSIG}.

We can add other primitives to Simplicity to access specific fields of transaction data in order to implement features such as Bitcoin's timelock operations~\cite{cltv,csv}.
While Simplicity avoids providing direct access to persistent storage, one could imagine an application where transactions contain data for transactional updates to a persistent store.
Simplicity could support primitives to read the details of these transactional updates and thereby enforce any set of programmable rules on them.

\subsection{Merklized Abstract Syntax Tree}

Recall that when using P2SH, users commit to their program by hashing it and placing that hash in the outputs of transactions.
Only when redeeming their funds does the user reveal their program, whose hash must match the committed hash.

\textit{Pay to Merklized Abstract Syntax Tree}, or \textit{P2MAST}~\cite{mast}, enhances the P2SH idea.
Instead of hashing a linear encoding of a Simplicity expression, we use the SHA-256's block compression function, $\compress : {\mathbb 2}^{256} \times {\mathbb 2}^{512} \rightarrow {\mathbb 2}^{256}$, to recursively hash Simplicity sub-expressions.
This computes a Merkle root of Simplicity's abstract syntax tree that we denote by $\mr{\cdot}$ and define as

\begin{align*}
 \mr{\iden} &\coloneqq \compress\tuple{\tg_{\iden}}{\repr{0}{{\mathbb 2}^{512}}}\\
 \mr{\comp s\,t} &\coloneqq \compress\tuple{\tg_{\comp}}{\tuple{\mr{s}}{\mr{t}}}\\
 \mr{\unit} &\coloneqq \compress\tuple{\tg_{\unit}}{\repr{0}{{\mathbb 2}^{512}}}\\
 \mr{\injl t} &\coloneqq \compress\tuple{\tg_{\injl}}{\tuple{\mr{t}}{\repr{0}{{\mathbb 2}^{256}}}}\\
 \mr{\injr t} &\coloneqq \compress\tuple{\tg_{\injr}}{\tuple{\mr{t}}{\repr{0}{{\mathbb 2}^{256}}}}\\
 \mr{\case s\,t} &\coloneqq \compress\tuple{\tg_{\case}}{\tuple{\mr{s}}{\mr{t}}}\\
 \mr{\pair s\,t} &\coloneqq \compress\tuple{\tg_{\pair}}{\tuple{\mr{s}}{\mr{t}}}\\
 \mr{\take t} &\coloneqq \compress\tuple{\tg_{\take}}{\tuple{\mr{t}}{\repr{0}{{\mathbb 2}^{256}}}}\\
 \mr{\drop t} &\coloneqq \compress\tuple{\tg_{\drop}}{\tuple{\mr{t}}{\repr{0}{{\mathbb 2}^{256}}}}\\
 \mr{\sigHash} &\coloneqq \compress\tuple{\tg_{\sigHash}}{\repr{0}{{\mathbb 2}^{512}}}\\
\end{align*}
where $\tg_c : {\mathbb 2}^{256}$ is an appropriate unique set of initial vectors per combinator. % for the SHA-256 block compression function.
We use the SHA-256 hash of the combinator name for its $\tg$ value.

When computing Merkle roots, like other kinds of static analysis, the intermediate results of sub-expressions can be shared.
Bitcoin Script cannot share sub-expressions in this manner due to the linear encoding of Bitcoin Script programs.

Another advantage of Merkle roots is that unused branches of $\case$ expressions can be pruned at redemption time.
Each unused branch can be replaced with the value of its Merkle root.
This saves on bandwidth, storage costs, and enhances privacy by not revealing more of a Simplicity program than necessary.
Again, this is something not possible in Bitcoin Script.

To enable redemption of pruned Simplicity expressions, we add two new combinators to replace the $\case$ combinator when pruning.

\[
\begin{array}{cc}
\inferrule {s : A \times C \vdash D \\ h : {\mathbb 2}^{256}} {\assertl s\, h : (A + B) \times C \vdash D}
&
\inferrule {h : {\mathbb 2}^{256} \\ t : B \times C \vdash D} {\assertr h\, t : (A + B) \times C \vdash D}
\end{array}
\]

These assertion combinators follow the same form as the $\case$ combinator, except one branch is replaced by a hash.
The semantics of these combinators require that the first component of their input be $\sigmaL$ or $\sigmaR$ as appropriate and then they evaluate the available branch.
The $h : {\mathbb 2}^{256}$ values do not affect the semantics; they instead influence the Merkle root computation.

\begin{align*}
 \mr{\assertl s\,h} &\coloneqq \compress\tuple{\tg_{\case}}{\tuple{\mr{s}}{h}}\\
 \mr{\assertr h\,t} &\coloneqq \compress\tuple{\tg_{\case}}{\tuple{h}{\mr{t}}}\\
\end{align*}

Notice that we use $\tg_{\case}$ for the tags of the assertions.
During redemption, one can replace $\case$ statements with appropriate assertions while still matching the committed Merkle root of the whole expression.

\[
\mr{\case s\,t} = \mr{\assertl s\,\mr{t}} = \mr{\assertr \mr{s}\,t}
\]
As long as the assertions hold, the computation remains unchanged.
If an assertion fails, because you pruned off a branch that was actually necessary, then the computation fails and a transaction trying to redeem funds in that manner is invalid.

Because failure is now a possible result during evaluation, we add a combinator $\fail$ to allow one to develop programs that use assertions.
\[
\inferrule {}{\fail : A \vdash B}
\]
\[
\mr{\fail} \coloneqq \compress\tuple{\tg_{\fail}}{\repr{0}{{\mathbb 2}^{512}}}
\]

When combined with the signature verification expression, one can build the equivalent of Bitcoin's \textsf{OP\_CHECKSIGVERIFY} operation to assert that a signature is valid. 

Degenerate assertions can also be used to enhance privacy by mixing entropy into Merkle roots of sub-expressions.
Given $t : A \vdash B$ the expression

\[
\comp\,(\pair\,(\injl \iden)\,\unit)\,(\assertl\,(\take t)\,h) : A \vdash B
\]
is semantically identical to $t$, but mixes $h$ into its Merkle root computation.
When used inside branches that are likely to be pruned, this prevents adversaries from effectively grinding out Simplicity expressions to see if they match the hash of the missing branch.

The Merkle root of an expression does not commit to its types.
We use first-order unification~\cite{unification} to perform type inference on Simplicity expressions, replacing any remaining type variables with the unit type.
Because the types of pruned branches are discarded, the inferred types may end up smaller than in the originally committed program.
When this happens, the memory requirements for evaluation with the Bit Machine may also decrease.

\subsection{Witness Values}

During redemption, the user must provide inputs, such as digital signatures and other witness data, in order to authorize a transaction.
Rather than passing this input as an argument to the Simplicity program, we embed such witness values directly into the Simplicity expressions using the $\witness$ combinator.

\[
\inferrule {b : B}{\witness b : A \vdash B}
\]

Semantically, the witness combinator just denotes a constant function, which is something we can already write in core Simplicity.
The difference lies in its Merkle root computation.

\[
 \mr{\witness b} \coloneqq \compress\tuple{\tg_{\witness}}{\repr{0}{{\mathbb 2}^{512}}}\\
\]

The value $b$ is not committed by the Merkle root.
This allows the user to set $\witness$ value at redemption time, without affecting the expression's Merkle root.
Users can use $\witness$ combinators at places where digital signatures are needed and at places where choices are made at redemption time.

Using $\witness$ combinators enhances privacy because they are pruned away when they occur in unused branches.
This leaves little evidence that there was an optional input at all.

The amount of witness data allowed at redemption time is limited at commitment time because each witness data type is finite and there can only be a finite number of witness combinators committed.
This limits the computational power of Simplicity.
For example, it is not possible to create a Simplicity program that allows hashing of an unbounded amount of witness data in order to validate a digital signature. % provided from an authorativive source.
However, it is possible to create a Simplicity program that allows a large, but bounded, amount of witness data to be hashed that is more than will ever be needed in practice.
Thanks to pruning, only the code used to hash the amount of witness data that actually occurs need be provided at redemption time.

Type inference determines the minimal type for witness values.
This prevents them from being filled with unnecessary data.

The $\witness$ combinator is not allowed to be used in jets.

\subsection{Extended Simplicity Semantics}

In order to accommodate these extensions to core Simplicity, we need to broaden Simplicity's semantics.
We use a monad, $\monad$, to capture the new effects of these extensions.

\[
\monad(A) \coloneqq \Transaction \rightarrow A_\bot
\]

This monad is a combination of an environment monad (also called a reader monad) with an exception monad.

This monad provides implicit access to a value of $\Transaction$ type, which we define to be the type of transaction data for the particular blockchain.
During evaluation, it is filled with the transaction containing the input redeeming the Simplicity program being evaluated.
This is used to give semantics to the $\sigHash$ and other similar primitives.

The monad also adds a failure value to the result type, denoted by $\bot : A_\bot$
This allows us to give semantics to assertions and $\fail$.

Given an expression $t : A \vdash B$, we denote this extended semantics by $\denoteM{t}: A \rightarrow \monad(B)$.
The core Simplicity semantics are lifted in the natural way so that

\[
\denoteM{t}(a) = \lambda e : {\Transaction}. \denote{t}(a)
\]
holds whenever $t$ is composed of only core Simplicity combinators.
The extended set of combinators have the following semantics

\begin{align*}
 \denoteM{\sigHash}(a) &\coloneqq \lambda e. \makeSigHash(a, e)\\
 \denoteM{\assertl s\,h}\tuple{\sigmaL(a)}{c} &\coloneqq \lambda e. \denoteM{s}\tuple{a}{c}(e)\\
 \denoteM{\assertl s\,h}\tuple{\sigmaR(b)}{c} &\coloneqq \lambda e. \bot\\
 \denoteM{\assertr h\,t}\tuple{\sigmaL(a)}{c} &\coloneqq \lambda e. \bot\\
 \denoteM{\assertr h\,t}\tuple{\sigmaR(b)}{c} &\coloneqq \lambda e. \denoteM{t}\tuple{b}{c}(e)\\
 \denoteM{\fail}(a) &\coloneqq \lambda e. \bot\\
 \denoteM{\witness b}(a) &\coloneqq \lambda e. b\\
\end{align*}
where $\makeSigHash(a, e)$ is a function that returns a hash of a given transaction $e$ in accordance with the given SigHash type $a$.

The operational semantics are also extended to support our new combinators by providing the Bit Machine with access to the transaction data and by adding an explicit $\mathbf{crash}$ instruction to support assertions and $\fail$.

We note that the effects captured by our monad are commutative and idempotent.\footnote{We use the terms `commutative' and `idempotent' in the sense of King and Wadler~\cite{King1993} as opposed to the traditional category theoretic definition of an idempotent monad.}
While Simplicity's operational semantics implicitly specify an order of evaluation, the extended denotational semantics are independent of evaluation order.
This simplifies formal reasoning about Simplicity programs that use the extended semantics.
That said, we expect the majority of Simplicity program's sub-expressions to be written in core Simplicity whose denotational semantics contain no effects at all, making reasoning about them simpler still.

\subsection{Using Simplicity Programs in Blockchains}

Simplicity programs are Simplicity expressions of type $p : {\mathbb 1} \vdash {\mathbb 1}$, that may use any of our extended combinators.
Users transact by constructing their Simplicity program and computing its Merkle root $\mr{p}$.
They have their counterparty create a transaction that sends funds to that hash.

Later, when a user wants to redeem their received funds, they create a transaction whose input contains a Simplicity program whose Merkle root matches the previous hash.
At this point they have the ability to set the witness values and prune any unneeded branches from $\case$ expressions.
The witness combinators handle the program's effective inputs, so the program $p$ is evaluated at the value $\splat : {\mathbb 1}$.
No output is needed because the program can use assertions to fail in case of invalid witnesses.
If the execution completes successfully, without failing, the transaction is considered valid.

The basic signature program that mimics Bitcoin's basic signature program is composed of the following core Simplicity expressions 

\begin{gather*}
 \checkSig : \Signature \times (\PubKey \times {\mathbb 2}^{256}) \vdash {\mathbb 2}\\
 \pubKey : A \vdash \PubKey\\
\end{gather*}
where $\checkSig$ checks whether a given signature validates for a given public key and message hash, and $\pubKey$ returns a user's specific public key.

These expressions can be combined into a basic signature verification program.

\begin{multline*}
\basicVerify b\,c \coloneqq \comp\,(\pair\,(\witness b)\\(\pair \pubKey\,(\comp\,(\witness c) \sigHash)))\\(\comp\,(\pair \checkSig \unit)\,(\case \fail \unit))
\end{multline*}

Other, more complex programs can be built to perform multi-signature checks, to implement sophisticated smart contracts such as zero-knowledge contingent payments, or to create novel smart contracts.

\section{Results and Future Work}

In many ways, Simplicity is best characterized by what features it leaves out rather than what features it contains.
\begin{itemize}
\item Simplicity has no state.
Purely functional, expression-based languages facilitate equational reasoning about the semantics of expressions.
For example, there are no concerns about aliased references to a global heap, so there is no need to work with separation logic or Hoare logic.
\item Simplicity has no named variables.
Using combinators lets us avoid dealing with binders and environments for bound variables.
This helps keep our interpreter and static analyses simple and further eases equational reasoning about Simplicity expressions.
\item Simplicity has no function types and therefore no higher-order functions.
While it is possible to compute upper bounds of computation resources of expressions in the presence of function types, it likely that those bounds would be so far above their actual needs that such analysis would not be useful.
\item Simplicity has no unbounded loops or recursion.
It is possible to build smart contracts with state carried through loops using covenants~\cite{oconnor2016}, without requiring unbounded loops within Simplicity itself.
Bounded loops, such as the 64 rounds needed by our SHA-256 implementation, can be achieved by unrolling the loop.
Because of sub-expression sharing, this doesn't unreasonably impact program size.
We do not directly write Simplicity, rather we use functions written in Coq or Haskell to generate Simplicity.
These languages do support recursion and we use loops in these meta-languages to generate unrolled loops in Simplicity.
\end{itemize}

Throughout this paper we have noted which theorems we have verified with Coq.
Other proofs are under development.
In particular, we plan to formally verify the denotational and operational semantics of the full Simplicity language, including assertions and blockchain primitives.

To validate the suitability of Simplicity, we will be building example smart contracts in a test blockchain or sidechain~\cite{sidechains} application using Simplicity.

We also plan to use the VST project~\cite{appel} to prove the correctness of our C implementation of the Bit Machine.
This would let us formally verify that the assembly code generated by the CompCert compiler~\cite{compcert} from our C implementation respects Simplicity's formal semantics.
In particular, we would be able to prove that substituting jets, such as SHA-256, with fast C or assembly implementations preserves the semantics of Simplicity.

Simplicity is designed as a low-level language interpreted by blockchain software.
We anticipate higher-level languages will be used for development and compiled to Simplicity.
Ivy~\cite{ivy} and the $\Sigma$-State Authentication Language~\cite{chepurnoy}, are existing efforts that may be suitable for being compiled to Simplicity.
If these higher-level languages come with formal semantics of their own, we will have the opportunity to prove correct the compiler to Simplicity for these languages.
For the time being, generating Simplicity with our Haskell and Coq libraries is possible.

\section{Conclusion}

We have defined a new language designed to be used for crypto-currencies and blockchains.
It could be deployed in new blockchain applications, including sidechains~\cite{sidechains}, or possibly in Bitcoin itself.
Simplicity has the potential to be used in any application where finitary programs need to be transported and executed under potentially adversarial conditions.

Our language is bounded, without loops, but is expressive enough to represent any finitary function.
These constraints allow for general purpose static analysis that can effectively bound the amount of computational resources needed by a Simplicity program prior to execution.

Simplicity has simple, functional semantics, which make formal reasoning with software proof assistants relatively easy.
This provides the means for people who develop smart contract to formally verify the correctness of their programs.
% While we cannot guarentee people will use this ability to produce correct smart contracts, we believe that for other languages without formal semantics we can pretty much guarentee people will continute to build incorrect smart contracts that will be exploited.

We have written several low-level functions in Simplicity such as addition, subtraction, and multiplication for various finite-bit words and formally verified their correctness.
With these we have built mid-level functions such as elliptic curve addition and multiplication, ECDSA signature validation, and a SHA-256 compression function.
This is already sufficient to create simple single and multi-signature verification programs in Simplicity.
We have formally verified our SHA-256 compression function is correct and have plans to formally verify the remaining functions.

\newpage\appendix
\section{Simplicity Semantics in Coq}\label{FormalDef}

Below is the formal definition of core Simplicity as expressed in Coq~\cite{coq}.
The \verb|Term| type is the type of well-typed Simplicity expressions.
The \verb|eval| function provides the denotational semantics of well-typed Simplicity expressions.

\verbatiminput{Simplicity.v}

\section{Full Word Adders in Simplicity}\label{FullAdder}

Below we recursively define Simplicity programs for ripple-carry full-adders for any $2^n$-bit word size.
A full-adder takes two $n$-bit words and a carry-in bit as inputs and returns a carry-out bit and an $n$-bit word.

\begin{align*}
 \fulladder_n &: ({\mathbb 2}^n \times {\mathbb 2}^n) \times {\mathbb 2} \vdash {\mathbb 2} \times {\mathbb 2}^n\\
 \fulladder_1 &\coloneqq \begin{aligned}[t]
 &\comp\,(\pair\,(\take \adder)\,(\drop \iden))\\
 &(\comp\,(\pair\,(\take\,(\take \iden))\\
 &(\comp\begin{aligned}[t]&(\pair\,(\take\,(\drop \iden))\,(\drop \iden))\\
 &\adder))\end{aligned}\\
 &(\pair\begin{aligned}[t]&(\case\,(\drop\,(\take \iden))\,(\injr \unit))\\
 &(\drop\,(\drop \iden))))
 \end{aligned}
 \end{aligned}
 \\
 \fulladder_{2n} &\coloneqq\begin{aligned}[t]
 &\comp\,(\pair\,(\take\,(\pair\begin{aligned}[t]&(\take\,(\take \iden))\\&(\drop\,(\take \iden))))\end{aligned}\\
 &(\comp\begin{aligned}[t]&(\pair\begin{aligned}[t]&(\take\,(\pair\begin{aligned}[t]&(\take\,(\drop \iden))\\&(\drop\,(\drop \iden))))\end{aligned}\\&(\drop \iden))\end{aligned}\\
  &\fulladder_n))\end{aligned}\\
 &(\comp\begin{aligned}[t]&(\pair\,(\drop\,(\drop \iden))\\
  &(\comp\begin{aligned}[t]&(\pair\begin{aligned}[t]&(\take \iden)\\&(\drop\,(\take \iden)))\end{aligned}\\
   &\fulladder_n))\end{aligned}\end{aligned}\\
 &(\pair\begin{aligned}[t]&(\drop\,(\take \iden))\\&(\pair\,(\drop\,(\drop \iden))\,(\take \iden))))
 \end{aligned}
 \end{aligned}
\end{align*}

\begin{theorem}[Full Adder Correct]
Let $n$ be a power of two.
Let $a, b : {\mathbb 2}^n$ be two $n$-bit words.
Let $c : {\mathbb 2}$ be a bit.
Then 
\[
\interp{x}{\mathbb 2} \cdot 2^n + \interp{y}{{\mathbb 2}^n} = \interp{a}{{\mathbb 2}^n} + \interp{b}{{\mathbb 2}^n} + \interp{c}{\mathbb 2}
\]
where
\[
\denote{\fulladder}\tuple{\tuple{a}{b}}{c} = \tuple{x}{y}
\]
\end{theorem}

We have verified the above theorem in Coq.

We have similarly defined multiplication for $2^n$-bit words and proven its correctness theorem in Coq.

\section{Tail Composition Optimization}\label{app:tco}

We define a variant of the translation of Simplicity to the Bit Machine, $\bm{\cdot}$, to add a tail composition optimization (TCO).
This optimization moves the $\mathbf{dropFrame}$ instruction earlier, potentially reducing the memory requirements for execution by the Bit Machine.
This is analogous to the tail call optimization found in other languages.
Our new definition will be a pair of mutually recursive functions, $\tcoff{\cdot}$ and $\tcon{\cdot}$.

The definition of $\tcoff{\cdot}$ is identical to that of $\bm{\cdot}$, replacing $\bm{\cdot}$ by $\tcoff{\cdot}$, except for the $\tcoff{\comp s\,t : A \vdash C}$ clause which is given below.

\begin{multline*}
 \tcoff{\comp s\,t : A \vdash C} \coloneqq
   \mathbf{newFrame}(\bitSize(B));\\\tcoff{s : A \vdash B};
   \mathbf{moveFrame};\tcon{t : B \vdash C}\\
\end{multline*}

In the tail position of the $\comp$ combinator, we removed the $\mathbf{dropFrame}$ instruction and call $\tcon{\cdot}$ instead.
The definition of $\tcon{\cdot}$ is given in Figure~\ref{tcondef}.

\begin{figure*}[t]
\begin{align*}
 \tcon{\iden : A \vdash A} &\coloneqq \mathbf{copy}(\bitSize(A));\mathbf{dropFrame}\\
 \tcon{\comp s\,t : A \vdash C} &\coloneqq
   \begin{aligned}[t]
   &\mathbf{newFrame}(\bitSize(B));\tcon{s : A \vdash B};\\
   &\mathbf{moveFrame};\tcon{t : B \vdash C}
   \end{aligned}\\
 \tcon{\unit : A \vdash {\mathbb 1}} &\coloneqq \mathbf{dropFrame}\\
 \tcon{\injl t : A \vdash B + C} &\coloneqq \mathbf{write}(\mathtt{0});\mathbf{skip}(\padl(B,C));\tcon{t : A \vdash B}\\
 \tcon{\injr t : A \vdash B + C} &\coloneqq \mathbf{write}(\mathtt{1});\mathbf{skip}(\padr(B,C));\tcon{t : A \vdash C}\\
 \tcon{\case s\,t : (A + B) \times C \vdash D} &\coloneqq \begin{aligned}[t]
   &\text{match }\mathbf{read}\text{ with}\\
   &\begin{cases}
   \texttt{0} \rightarrow \begin{aligned}[t]
                          &\mathbf{fwd}(1 + \padl(A,B));\\
                          &\tcon{s : A \times C \vdash D}
                          \end{aligned}\\
   \texttt{1} \rightarrow \begin{aligned}[t]
                          &\mathbf{fwd}(1 + \padr(A,B));\\
                          &\tcon{t : B \times C \vdash D}
                          \end{aligned}
   \end{cases}
   \end{aligned}\\
 \tcon{\pair s\,t : A \vdash B \times C} &\coloneqq \tcoff{s : A \vdash B};\tcon{t : A \vdash C}\\
 \tcon{\take t : A \times B \vdash C} &\coloneqq \tcon{t : A \vdash C}\\
 \tcon{\drop t : A \times B \vdash C} &\coloneqq \mathbf{fwd}(\bitSize(A));\tcon{t : B \vdash C}\\
\end{align*}
\caption{Operational semantics for Simplicity using the Bit Machine with TCO.}\label{tcondef}
\end{figure*}

We have formally verified in Coq the following correctness theorem for $\tcoff{\cdot}$.

\begin{theorem}[Correctness of TCO Operational Semantics]\label{ctcoos}
Let $t : A \vdash B$ be any Simplicity expression.
Let $v : A$ be any value of type $A$.
Initialize a Bit Machine with 
\begin{itemize}
\item the value $\cells{v}{A}$ as the only frame on the read-frame stack with its cursor at the beginning of the frame, and
\item the value $\mathtt{[?]}^{\bitSize(B)}$ as the only frame on the write-frame stack with its cursor at the beginning of the frame.
\end{itemize}
After executing the instructions $\tcoff{t : A \vdash B}$, the final state of the Bit Machine has
\begin{itemize}
\item the value $\cells{v}{A}$ as the only frame on the read-frame stack with its cursor at the beginning of the frame, and
\item the value $\cells{\denote{t}(v)}{B}$ as the only frame on the write-frame stack with its cursor at the end of the frame.
\end{itemize}
In particular, the Bit Machine never crashes during this execution.
\end{theorem}

The proof proceeds by establishing that the machine state transformation induced by executing the instructions 
\[\tcoff{t : A \vdash B};\mathbf{dropFrame}\]
and the instructions 
\[\tcon{t : A \vdash B}\]
are identical.

We would like an improved static analysis of the memory use of this TCO execution.
Figure~\ref{cbtcodef} defines the static analysis $\cbtco(t : A \vdash B)$ which bounds the memory use of the TCO execution.
The static analysis becomes somewhat more intricate with the more complex translation.
But we may proceed with confidence because we have a formal verification in Coq of the following theorem.

\begin{figure*}[h]
\begin{align*}
 \ecbtco(\iden : A \vdash A) &\coloneqq \tuple{0}{0}\\
 \ecbtco(\comp s\,t : A \vdash C) &\coloneqq \begin{aligned}[t]
   &\text{let }\tuple{n_1}{n_2} =\\&\ecbtco(s : A \vdash B)\text{ in}\\
   &\text{let }\tuple{m_1}{m_2} =\\&\ecbtco(t : B \vdash C)\text{ in}\\
   &\text{let }b = \bitSize(B)\text{ in}\\
   &\tuple{\max(b + n_1, m_1, b + m_2)}{b + n_2}
   \end{aligned}\\
 \ecbtco(\unit : A \vdash {\mathbb 1)} &\coloneqq \tuple{0}{0}\\
 \ecbtco(\injl t : A \vdash B + C) &\coloneqq \ecbtco(t : A \vdash B)\\
 \ecbtco(\injr t : A \vdash B + C) &\coloneqq \ecbtco(t : A \vdash C)\\
 \ecbtco(\case s\,t : (A + B) \times C \vdash D) &\coloneqq \begin{aligned}[t]
   &\text{let }\tuple{n_1}{n_2} =\\&\ecbtco(s : A \times C \vdash D)\\&\text{ in}\\
   &\text{let }\tuple{m_1}{m_2} =\\&\ecbtco(t : B \times C \vdash D)\\&\text{ in}\\
   &\tuple{\max(n_1, m_1)}{\max(n_2, m_2)}
   \end{aligned}\\
 \ecbtco(\pair s\,t : A \vdash B \times C) &\coloneqq \begin{aligned}[t]
   &\text{let }\tuple{n_1}{n_2} =\\&\ecbtco(s : A \vdash B)\text{ in}\\
   &\text{let }\tuple{m_1}{m_2} =\\&\ecbtco(t : A \vdash C)\text{ in}\\
   &\tuple{m_1}{\max(n_1, n_2, m_2)}
   \end{aligned}\\
 \ecbtco(\take t : A \times B \vdash C) &\coloneqq \ecbtco(t : A \vdash C)\\
 \ecbtco(\drop t : A \times B \vdash C) &\coloneqq \ecbtco(t : B \vdash C)\\
\end{align*}
\[
 \cbtco(t : A \vdash B) \coloneqq \begin{aligned}[t]
   &\text{let }\tuple{n_1}{n_2} = \ecbtco(t : A \vdash B)\text{ in}\\
   &\bitSize(A) + \bitSize(B) + \max(n_1, n_2)
   \end{aligned}\\
\]
\caption{Definition of $\cbtco$.}\label{cbtcodef}
\end{figure*}

\begin{theorem}[Static Analysis of Cell Usage with TCO]
Let $t : A \vdash B$ be any Simplicity expression.
Let $v : A$ be any value of type $A$.
Initialize a Bit Machine as specified in Theorem~\ref{ctcoos}.
The maximum number of cells in both stacks of the Bit Machine at any point during the execution of $\tcoff{t : A \vdash B}$ from this initial state never exceeds $\cbtco(t : A \vdash B)$.
\end{theorem}

The proof proceeds by establishing that the additional memory used by $\tcoff{t : A \vdash B}$ and $\tcon{t : A \vdash B}$ in any machine state is no more than $\max(n_1, n_2)$ and $\max(n_1 - m, n_2)$ respectively where $\tuple{n_1}{n_2} = \ecbtco(t : A \vdash B)$ and $m$ is the number of cells in the active read frame before executing $\tcon{t : A \vdash B}$.

There is a possibility for a head composition optimization where the $\mathbf{newFrame(n)}$ instruction is delayed in order to potentially save memory.
It is unclear to us if this is worth the added complexity, so we have not pursued this yet.

\section*{Acknowledgements}
Thank you to Shannon Appelcline for his help editing this paper.
\FloatBarrier
\bibliographystyle{plain}
\bibliography{Simplicity}

\end{document}